\documentclass {elsart}
\usepackage{graphicx}

\usepackage{graphics}
\usepackage{graphicx}
\usepackage{epsfig}

\usepackage{amssymb}

\begin{document}

\begin{frontmatter}



\title{Legendre transform structure and extremal properties of the relative Fisher information}


\author[SRC]{R. C. Venkatesan\corauthref{cor}}
\corauth[cor]{Corresponding author.}
\ead{ravi@systemsresearchcorp.com}
\author[UNLP]{A. Plastino}
\ead{plastino@fisica.unlp.edu.ar}

\address[SRC]{Systems Research Corporation,
Aundh, Pune 411007, India}
\address[UNLP]{IFLP, National University La Plata \&
National Research Council (CONICET)\\ C. C., 727 1900, La Plata,
Argentina}

\begin{abstract}
Variational extremization of the relative Fisher information (RFI, hereafter) is performed.   Reciprocity relations, akin to those of
thermodynamics are derived, employing the extremal results of the  RFI  expressed in terms of probability
amplitudes. A time independent Schr\"{o}dinger-like equation  (Schr\"{o}dinger-like link) for the RFI is derived.  The concomitant Legendre transform
structure (LTS, hereafter) is developed by utilizing a generalized RFI-Euler
theorem, which shows  that the entire mathematical structure of thermodynamics
translates into the RFI framework,  both for equilibrium and
non-equilibrium cases.  The qualitatively distinct nature of the present
results  \textit{vis-\'{a}-vis} those of  prior studies utilizing
the Shannon entropy and/or the Fisher information measure (FIM, hereafter) is
 discussed.   A principled relationship between the RFI and the FIM frameworks is derived.  The utility of this relationship is demonstrated by an example wherein the energy eigenvalues of the Schr\"{o}dinger-like link for the RFI is inferred solely using the quantum mechanical virial theorem and the LTS of the RFI.
\end{abstract}

\end{frontmatter}

\section{Introduction}
The Fisher information measure (FIM, hereafter) [1-3]
\begin{equation}
I[f] =
\int_{\Re ^n } {f\left( \textbf{x} \right)} \left( {\frac{{\partial \ln f\left( \textbf{x} \right)}}{{\partial \textbf{x}}}} \right)^2 d\textbf{x} = \int_{\Re ^n } {\frac{1}{{f\left( \textbf{x} \right)}}} \left( {\frac{{\partial f\left( \textbf{x} \right)}}{{\partial \textbf{x}}}} \right)^2 d\textbf{x},
\end{equation}
 where $\textbf{x}$ is a vector ($\frac{{\partial f\left( {x_i } \right)}}{{\partial x_j }} = 0;i \ne j$), has played a prominent role in statistics, information theory, physics, and allied disciplines.  In addition to the applications cited in Ref. [3],
 the FIM has also been successfully employed in areas as diverse as biology, social science, econophysics, and encryption of covert information
 amongst a number of other applications (eg. see [4]).   The relative Fisher information (RFI, hereafter) defined by [5, 6]
\begin{equation}
\Im \left[f|g \right] =
\int_{\Re ^n } {f\left( \textbf{x} \right)\left| {\nabla \ln \frac{{f\left( \textbf{x} \right)}}{{g\left( \textbf{x} \right)}}} \right|} ^2 d\textbf{x},
\end{equation}
where $\left|  \bullet  \right|^2$ is the square norm, has been primarily studied within the context of mathematical physics and optimal
transportation in statistical physics (eg. Refs. [7,8] and the references therein).  Note that $\Im [f|g]=0$ when $f(\textbf{x})=g(\textbf{x})$, and $\Im[f|g] \neq \Im[g|f]$ (asymmetric).  An alternate form of the $n$-dimensional RFI has been suggested by Carlen and Soffer [9] for a Gaussian $g(x)$
\begin{equation}
\Im \left[ f|g \right] =
4\int_{\Re ^n } {\left| {\left( {\nabla  + \frac{x}{2}} \right)\sqrt {f\left( x \right)} } \right|} ^2 d^n x,
\end{equation}
where ${\left( {\nabla  + \frac{x}{2}} \right)}$ is the \textit{lowering operator} of the harmonic oscillator Hamiltonian with ground state ${\sqrt
{g\left( x \right)} }$.  Here,  $g\left( x \right) = \left( {2\pi } \right)^{ - \frac{n}{2}} \exp \left[ { - \frac{{x^2 }}{2}} \right]$.

Recently, the RFI has been the subject of intense investigations in a quest to obtain a better perspective of its physical implications [10-14], and
to formally establish its role in information theory and estimation theory [15, 16].   Still, many of the fundamental properties and physical
implications of the RFI remain uninvestigated.  The RFI has recently been related to the Kullback-Leibler
divergence (K-Ld, hereafter)[2]
\begin{equation}
D\left[ f\|g \right] =
\int_{\Re ^n } {f\left( \textbf{x} \right)\ln \frac{{f\left( \textbf{x} \right)}}{{g\left( \textbf{x} \right)}}} d\textbf{x},
\end{equation}
with the aid of the de Bruijn identity [2] by  Verd\'{u} [15], and Guo, Shamai (Shitz), and Verd\'{u} [16] as
\begin{equation}
\frac{d}{{d\delta }}D\left[ {X + \sqrt \delta  Z\left\| {Y + \sqrt \delta  Z} \right.} \right]_{\delta  = 0}  =  - \frac{1}{2}\Im \left[ {X\left| Y
\right.} \right],
\end{equation}
where $X$ and $Y$ are random variables, and $Z$ is random variable (not necessarily Gaussian), which is independent of $X$ and $Y$.  When random
variables $X$ and $Y$ have the densities $f$ and $g$, respectively,
the K-Ld and the RFI of $X$ with respect to $Y$ are defined by
$D[X \| Y ] = D[f\|g]$ and $\Im[X|Y]=\Im[f|g]$, respectively.

Akin to the K-Ld, the RFI may not only be construed as being a
measure of uncertainty, but also a measure of discrepancy between
two probability densities.  The RFI relates to the FIM in a
similar manner to which the K-Ld relates to the Shannon entropy.
In contrast to the FIM and the RFI, whose derivative term induces
the effect of "localization", the Shannon entropy and the K-Ld are
"coarse-grained".

It is important to state that this Letter treats the case of one-dimensional time independent probability density functions and their concomitant probability amplitudes.  Thus, within the framework of optimal transportation theory [5], the analysis presented herein is applicable to steady-state models. On the other hand, the results of this paper are directly applicable to the RFI models studied in [13-16].  Further, the \emph{reference probability} $g(\textbf{x})$ in the expression of the RFI (Eq.(2)) may be treated as representing \emph{prior knowledge}.  Thus, like the K-Ld the RFI possesses the ability of being employed in inference studies.

This Letter accomplishes the following objectives: $(i)$ Setting $g(x)=\exp[-V(x)]$ where $V(x)$ is a convex potential described in Section 2, a principled relation between the RFI, the FIM, and derivative terms of the convex potential $V(x)$ is established (Section 3). $(ii)$  A time independent
Schr\"{o}dinger-like Sturm-Liouville equation (hereafter referred to as the Schr\"{o}dinger-like link for the RFI), resulting from the variational
extremization of the RFI is derived (Section 4). $(iii)$  The reciprocity relations and the Legendre
transform structure for the RFI are derived, thereby explicitly
demonstrating that the entire mathematical structure of thermodynamics
translates into the RFI framework (Sections 5 $\&$ 6). $(iv)$  The utility of the relationship between the RFI and FIM frameworks (derived in Section 3) is demonstrated by inference of the energy eigenvalues of the Schr\"{o}dinger-like link for the RFI (Section 7).  This is achieved by solely utilizing the quantum mechanical virial theorem [17] and the Legendre transform structure of the RFI, without recourse to solving the Schr\"{o}dinger-like link for the RFI. To the best of the authors' knowledge, none of the above stated results have been hitherto accomplished.

\section{Theoretical preliminaries}
It is common in optimal transportation theory (eg., see [5, 7, 8]) to define the probability $g(\textbf{x})$ as a \textit{reference probability}, alternately
referred to as the \textit{equilibrium probability}.  Specializing Refs. [5, 7, 8] and numerous other works to the one-dimensional case
\begin{equation}
\begin{array}{l}
 g\left( x \right) = e^{ - V\left( x \right)} , \\
\\
\int {e^{ - V\left( x \right)} dx}  = 1, \\
\end{array}
\end{equation}
which is known as the \textit{Gibbs form}, where $V(x)$ is a convex potential.Thus, (2) is re-written with the aid of (6) as
\begin{equation}
\Im \left[ {f\left| {e^{ - V\left( x \right)} } \right.} \right] = \int {f\left( x \right)\left| {\nabla \left( {\ln f\left( x \right) + V\left( x
\right)} \right)} \right|^2 dx}.
\end{equation}
It is noteworthy to mention that the term $\ln f(x) + V(x)$ in (7) is exactly the potential $\Psi$ whose expectation in time independent stochastic thermodynamics is an analog to the Helmholtz free energy, for $k_BT=constant$ [18, 19].  Note that in this Letter, all expectation values denoted by $
\left\langle  \bullet  \right\rangle$ are evaluated with respect to $f(x)=\psi^2(x)$.  Here, $\psi(x)$ is the probability amplitude which extremizes the RFI.

\section{Relation between the FIM and the RFI}
Expanding (7) yields
\begin{equation}
\begin{array}{l}
 \Im \left[ {f\left| {e^{ - V\left( x \right)} } \right.} \right] = \int {\left\{ {\frac{1}{{f\left( x \right)}}\left( {\frac{{df\left( x \right)}}{{dx}}} \right)^2  + 2V_x \left( x \right)\frac{{df\left( x \right)}}{{dx}} + V_x^2 \left( x \right)f\left( x \right)} \right\}} dx \\
 \\
  = \int {\frac{1}{{f\left( x \right)}}\left( {\frac{{df\left( x \right)}}{{dx}}} \right)^2 dx}  + \int {2V_x \left( x \right)\frac{{df\left( x \right)}}{{dx}}dx}  + \int {V_x^2 \left( x \right)f\left( x \right)} dx, \\
 \end{array}
\end{equation}
where $
V_x \left( x \right) = \frac{{dV\left( x \right)}}{{dx}}$.
Integrating by parts the second term in the second expression of (8), and specifying $f(x) V_{x}(x)$ to vanish at the boundaries yields
\begin{equation}
2\int {V_x \left( x \right)} \frac{{df\left( x \right)}}{{dx}}dx =  - 2\int {V_{xx} \left( x \right)} f\left( x \right)dx =  - 2\left\langle {V_{xx} \left( x \right)} \right\rangle,
\end{equation}
$V_{xx} \left( x \right) = \frac{{d^2 V\left( x \right)}}{{dx^2 }}$.  Substituting (9) into (8) yields
\begin{equation}
\Im \left[ {\left. f \right|e^{ - V\left( x \right)} } \right] = \int {\frac{1}{{f\left( x \right)}}} \left( {\frac{{df\left( x \right)}}{{dx}}} \right)^2 dx - 2\left\langle {V_{xx} \left( x \right)} \right\rangle  + \left\langle {V_x^2 \left( x \right)} \right\rangle.
\end{equation}
Invoking the definition of the FIM in (1) results in the \emph{critical} relationship
\begin{equation}
\Im \left[ {f\left| {e^{ - V\left( x \right)} } \right.} \right] = I\left[ f \right] - 2\left\langle {V_{xx} \left( x \right)} \right\rangle  + \left\langle {V_x^2 \left( x \right)} \right\rangle.
\end{equation}
\emph{Here, (11) unambiguously relates the RFI with the FIM and the expectations of the derivatives of the convex potential $V(x)$}.  Thus, (11) tacitly demonstrates that the results presented in this Letter for the RFI qualitatively differ from any results obtained for the FIM (eg. see [3]).

\section{Schr\"{o}dinger-like link for the RFI}
In physics, it is often desirable to express probabilities in the
form of amplitudes.   In the one-dimensional case, it is tenable
to treat probability amplitudes as real quantities [20].  Expressing (7) in terms of probability amplitudes by specifying $f(x)=\psi^2(x)$ and performing variational extremization, results in
\begin{equation}
\begin{array}{l}
\frac{\delta }{{\delta \psi \left( x \right)}}
\int {\left\{ {4\left( {\frac{{d\psi \left( x \right)}}{{dx}}} \right)} \right.} ^2  - 2V_{xx} \left( x \right)\psi^2(x)  +  V_x^2 \left( x \right)\psi^2(x)  \\
 \\
 \left. { - \sum\limits_{i = 1}^M {\lambda _i A_i \left( x \right)\psi ^2 \left( x \right)}  - \lambda _0 \psi ^2 \left( x \right)} \right\}dx = 0, \\
 \end{array}
\end{equation}
where $\left\langle {A_i \left( x \right)} \right\rangle  = \int {A_i \left( x \right)\psi ^2 \left( x \right)dx}$ are the constraint terms entailing $M$ Lagrange multipliers $\lambda_i$, and $
\int {\psi ^2 \left( x \right)dx}  = 1$ is the normalization condition. Carrying through with the variational extremization of (12) yields a time independent
Schr\"{o}dinger-like equation
\begin{equation}
 - \frac{1}{2}\frac{{d^2 \psi \left( x \right)}}{{dx^2 }} - U_{RFI}\left( x \right)\psi \left( x \right) =   \frac{{\lambda _0 }}{8}\psi \left( x \right).
\end{equation}
The \textit{pseudo-potential} in (13) comprising of data driven terms, and derivatives of the potential $V(x)$ is defined by
\begin{equation}
U_{RFI}\left( x \right) = \frac{1}{8}\left[ \sum\limits_{i = 1}^M {\lambda _i A_i \left( x \right)}   -  {V^2_x \left( x \right)}   + 2V_{xx} \left( x \right) \right].
\end{equation}
Note that (13) is a special case of the Sturm-Liouville equation denoted by:  $
 - \frac{d}{{dx}}\left[ {q\left( x \right)\frac{{d\psi \left( x \right)}}{{dx}}} \right] + g\left( x \right)\psi \left( x \right) = \mu h\left( x \right)\psi \left( x \right)$, where $q(x)$, $g(x)$ and $h(x)$ are suitable arbitrary functions, and $\mu$ is the eigenvalue.  Specifying: $
q\left( x \right) = 1,g\left( x \right) =  - U_{RFI} \left( x \right),h(x)=1$ and $
\mu  = \frac{{\lambda _0 }}{8} = E$ yields the form of the usual time independent Schr\"{o}dinger equation, for $
\frac{{\hbar ^2 }}{m} = 1$ and having energy eigenvalue $E$.

At this stage it is important to highlight two facts.
First, the normalization Lagrange
multiplier:$\lambda_0=\lambda_0(\lambda_1,...,\lambda_M)$, and, $
\psi \left( {x,\overrightarrow \lambda  } \right)$ are the
solution of (13), where $\overrightarrow \lambda$ is a M-vector of
Lagrange multipliers.  Next, the convex potential $V(x)$ and its
derivatives are not  functions of $\overrightarrow \lambda$, and
the exact form of $V(x)$ is assumed to be known \emph{a-priori}. For the sake of generality, the form of $V(x)$
is kept arbitrary in the above analysis. Note that $\psi(x)$ is the probability amplitude that extremizes the RFI.  It is interesting to note
that setting $g(x)=exp[-\sqrt{k}x^2]$ results in the potential for
the harmonic oscillator, which is one of the most fundamental
systems in quantum mechanics.

\section{Reciprocity relations}
It is known that standard thermodynamics makes use of derivatives
of the Shannon entropy with respect to both parameters, i.e. the Lagrange multipliers $\lambda_i$ and expectation values
$\left\langle {A_i \left( x \right)} \right\rangle $  (for instance, pressure and volume,
respectively). The basis for the reciprocity relations is the
generalized Euler theorem [21]. In
[22], it was shown that the Euler theorem [21] is recovered within
the Fisher-context by the formulation of a generalized
Fisher-Euler theorem.   Given a generic measure of uncertainty
$\aleph$ and expectation values $\left\langle {A_i \left( x \right)} \right\rangle$, the generalized Euler
theorem is required to be of the form [21]
\begin{equation}
\frac{{\partial \aleph
}}{{\partial \lambda _i }} = \sum\limits_{j = 1}^M {\lambda _j
\frac{{\partial \left\langle {A_j(x) } \right\rangle }}{{\partial
\lambda _i }}}.
\end{equation}
Here, $\aleph$ may be the Shannon entropy, the
FIM, or the RFI (as will be demonstrated in this Section).  Specifically, this Section qualitatively extends the
analysis in [22] by establishing a principled generalized
RFI-Euler theorem.
Substituting (14) into (13) and multiplying the resulting expression throughout by 8 yields
\begin{equation}
  - 4\frac{{d^2 \psi \left( x \right)}}{{dx^2 }} - \sum\limits_{i = 1}^M {\lambda _i A_i \left( x \right)\psi \left( x \right)}  - \left[ {2V_{xx} \left( x \right) - V_x^2 \left( x \right)} \right]\psi \left( x \right) = \lambda _0 \psi \left( x \right).
\end{equation}
From [3], the FIM after a single integration by parts is defined by
\begin{equation}
I\left[ \psi  \right] = 4\int {\left( {\frac{{d\psi \left( x \right)}}{{dx}}} \right)^2 dx =  - 4\int {\psi \left( x \right)\frac{{d^2 \psi \left( x \right)}}{{dx^2 }}dx} }.
\end{equation}
Multiplying (16) by $\psi(x)$, integrating and re-arranging the terms results in
\begin{equation}
 - 4\int {\psi \left( x \right)} \frac{{d^2 \psi \left( x \right)}}{{dx^2 }}dx - 2\left\langle {V_{xx} \left( x \right)} \right\rangle  + \left\langle {V_x^2 \left( x \right)} \right\rangle  = \lambda _0  +
\sum\limits_{i = 1}^M {\lambda _i \left\langle {A_i \left( x \right)} \right\rangle },
\end{equation}
where the normalization condition $\int {\psi ^2 \left( x \right)dx}  = 1$ has been invoked.  From (17) and the relation (11) it is readily seen that the LHS of (17) is a re-statement of the RHS of (11) expressed in terms of probability amplitudes, i.e.
\begin{equation}
\Im \left[ {\psi\left| {e^{ - V\left( x \right)/2} } \right.} \right] = I\left[ \psi \right] - 2\left\langle {V_{xx} \left( x \right)} \right\rangle  + \left\langle {V_x^2 \left( x \right)} \right\rangle.
\end{equation}
Thus, from (18) and (19) the following relation is obtained
\begin{equation}
\Im \left[ {\psi \left| {e^{ - V\left( x \right)/2} } \right.} \right] = \lambda _0  + \sum\limits_{i = 1}^M {\lambda _i \left\langle {A_i \left( x \right)} \right\rangle },
\end{equation}
where $\left\langle {A_i \left( x \right)} \right\rangle  = \int {A_i \left( x \right)\psi ^2 \left( x \right)dx}$.  Taking the derivative of (20) with respect to $\lambda_i$ yields
\begin{equation}
\frac{{\partial \Im \left[ {\psi \left| {e^{ - V\left( x \right)/2} } \right.} \right]}}{{\partial \lambda _i }} = \frac{{\partial \lambda _0 }}{{\partial \lambda _i }} + \left\langle {A_i(x) } \right\rangle  + \sum\limits_{\scriptstyle j = 1 \hfill \atop
  \scriptstyle j \ne i \hfill}^M {\lambda _j \frac{{\partial \left\langle {A_j(x) } \right\rangle }}{{\partial \lambda _i }}}.
\end{equation}
Specifying
\begin{equation}
\frac{{\partial \lambda _0 }}{{\partial \lambda _i }} =  - \left\langle {A_i(x) } \right\rangle,
\end{equation}
and substituting (22) into (21) yields
\begin{equation}
\frac{{\partial \Im \left[ {\psi \left| {e^{ - V\left( x \right)/2} } \right.} \right]}}{{\partial \lambda _i }} = \sum\limits_{j = 1}^M {\lambda _j \frac{{\partial \left\langle {A_j(x) } \right\rangle }}{{\partial \lambda _i }}}.
\end{equation}

  To demonstrate that the Lagrange multipliers and the expectation values are
   conjugate variables (see Section 6), it is required to establish: $\frac{{\partial \Im \left[ {\psi \left| {e^{ - V\left( x \right)/2} } \right.} \right]  }}{{\partial \left\langle {A_l(x) } \right\rangle }} = \lambda _l; l=i,j$.  While this is evident from (27) in Section 6, the thermodynamic counterpart of the generalized RFI-Euler theorem (23)
   is considered by evaluating the derivative of the RFI with respect to the expectation values as
\begin{equation}
\sum\limits_{i = 1}^M {\frac{{\partial \Im \left[ {\psi \left| {e^{ - V\left( x \right)/2} } \right.} \right] }}{{\partial \lambda _i }}\frac{{\partial \lambda _i }}{{\partial \left\langle {A_j } \right\rangle }}}
  = \sum\limits_{i = 1}^M {\sum\limits_{k = 1}^M {\lambda _k \frac{{\partial \left\langle {A_k(x) } \right\rangle }}{{\partial \lambda _i }}
  \frac{{\partial \lambda _i }}{{\partial \left\langle {A_j(x) } \right\rangle }}.} }
\end{equation}
Eq. (24) readily reduces to
\begin{equation}
\frac{{\partial \Im \left[ {\psi \left| {e^{ - V\left( x \right)/2} } \right.} \right]  }}{{\partial \left\langle {A_j(x) } \right\rangle }} = \lambda _j,~and ~ likewise,~\frac{{\partial \Im \left[ {\psi \left| {e^{ - V\left( x \right)/2} } \right.} \right]  }}{{\partial \left\langle {A_i(x) } \right\rangle }} = \lambda _i.
\end{equation}

The generalized Euler theorem for the Shannon entropy bears similarities to
that for the FIM, which has now been shown to bear similarities to that for the RFI.
 The utility of the generalized Euler theorem is to establish the conjugate relationship between the Lagrange
 multipliers and the expectation values for a given measure of uncertainty.
   To establish this conjugate relationship between the
   Lagrange multipliers and the expectation values for the RFI, it is
   imperative to utilize Eqs. (23) and (25) above.  For the case
    of the RFI, this conjugate relationship is established in Section 5.
    In summary, while (23) naturally bears similarities to the form of the generalized Fisher-Euler
theorem established in [22] and the Euler theorem within the Boltzmann-Gibbs-Shannon (B-G-S, hereafter)
 framework [21], it qualitatively differs
from the previous studies.

These qualitative distinctions may be summarized as  $(i)$ the
measure of uncertainty (which in this case is also a measure of
discrepancy) is  the RFI, $(ii)$ the expectations are evaluated
with respect to the probability amplitude $\psi(x)$, which extremizes the RFI,
and $(iii)$ the concomitant Lagrange multipliers, which comprise
the solution of (13), viz. $\psi(x,\overrightarrow\lambda)$,
are dissimilar to those obtained in previous studies. viz.
[21, 22]. Specifically, correspondence relationships between the
solutions and the Lagrange multipliers of the B-G-S model and the
FIM model have been established (eg., see [23])), and the
solutions of the variational extremizations of the Shannon entropy
and the FIM are known to coincide for the case of equilibrium
distributions [3].
Apart from seamlessly relating the RFI and the FIM frameworks, Eq. (11) also provides the basis for relating their respective extremal solutions (and by extension to the extremal solutions of the B-G-S framework) by drawing analogies to the approach adopted in Ref.
[23].  This is the objective of ongoing work.

\section{Legendre transform structure}
The objective of obtaining the Legendre transform structure is to
place the Lagrange multipliers ($\lambda_i$'s) and the expectation
values ($
\left\langle {A_i \left( x \right)} \right\rangle$'s) on an equal footing for the purpose of
determining $ {\Im \left[ {\psi \left| {e^{ - V\left( x \right)/2}
} \right.} \right]}$.   This entails that the Lagrange multipliers
and the expectation values play \textit{reciprocal and symmetric}
roles thermodynamically, thereby allowing for the input
information to be provided also in the form of the Lagrange
multipliers. In the usual case, of course, one employs expectation
values [24].  Re-stating (20) and re-arranging the terms
yields
\begin{equation}
\lambda _0= {\Im \left[ {\psi \left| {e^{ - V\left( x \right)/2} } \right.} \right]} - \sum\limits_{i = 1}^M {\lambda _i \left\langle {A_i(x) } \right\rangle }.
\end{equation}
From the arguments stated in Section 4,
$\Im=\Im(x,\overrightarrow\lambda)$.  From (26), it is evident
that
\begin{equation}
\lambda _0 \left( {\lambda _1 ,...,\lambda _M } \right) = \Im \left( {\left\langle {A_1(x) } \right\rangle ,...,\left\langle {A_M(x) } \right\rangle }\right) - \sum\limits_{i = 1}^M {\lambda _i \left\langle {A_i(x) } \right\rangle }.
\end{equation}
Here, (27) is the Legendre transform of the RFI since it
changes the identity of the relevant variables; viz.
$\Im(x,\overrightarrow\lambda)\leftrightarrow \Im \left(
{\left\langle {A_1(x) } \right\rangle ,...,\left\langle {A_M(x) }
\right\rangle } \right)$. Specifically, the Legendre transform
relates the input parameters (independent variables of the RFI), such that  $\left\{
{\Im ,\left\langle {A_1(x) } \right\rangle ,...,\left\langle {A_M(x) }
\right\rangle } \right\} \leftrightarrow \left\{ {\lambda _0 ,\vec
\lambda } \right\}$.  Note that within the context of (27),
$\lambda_0$ may also be construed as being the generalized
thermodynamic potential of the RFI.

From (27) and (25) (which has been derived on the basis of the generalized RFI-Euler theorem (23)), the following result is obtained
\begin{equation}
\frac{{\partial \lambda _0 }}{{\partial \lambda _i }} = \sum\limits_{j = 1}^M {\frac{{\partial \Im }}{{\partial \left\langle {A_j(x) } \right\rangle }}}
\frac{{\partial \left\langle {A_j(x) } \right\rangle }}{{\partial \lambda _i }} - \sum\limits_{j = 1}^M {\lambda _j } \frac{{\partial \left\langle {A_j(x)
}
\right\rangle }}{{\partial \lambda _i }} - \left\langle {A_i(x) } \right\rangle= - \left\langle {A_i(x) } \right\rangle.
\end{equation}
Eq.'s (25) and (28) yield
\begin{equation}
 \frac{{\partial \lambda _i }}{{\partial \left\langle {A_j(x) } \right\rangle }} = \frac{{\partial \lambda _j }}{{\partial \left\langle {A_i(x) }
 \right\rangle }} = \frac{{\partial ^2 \Im }}{{\partial \left\langle {A_i(x) } \right\rangle \partial \left\langle {A_j(x) } \right\rangle }}, \\
\end{equation}
and
\begin{equation}
 \frac{{\partial \left\langle {A_j(x) } \right\rangle }}{{\partial \lambda _i }} = \frac{{\partial \left\langle {A_i(x) } \right\rangle }}{{\partial
 \lambda
 _j }} =  - \frac{{\partial ^2 \lambda _0 }}{{\partial \left\langle {A_i(x) } \right\rangle \partial \left\langle {A_j(x) } \right\rangle }}, \\
\end{equation}
respectively. As a consequence of (30), the generalized RFI-Euler theorem (23) is re-stated as
\begin{equation}
\frac{{\partial \Im \left[ {\psi \left| {e^{ - V\left( x \right)/2} } \right.} \right]}}{{\partial \lambda _i }} = \sum\limits_{j = 1}^M {\lambda _j \frac{{\partial \left\langle {A_i(x) } \right\rangle }}{{\partial \lambda _j }}}.
\end{equation}

Here, Eqs. (25), (27) and (28)-(31) constitute the Legendre transform structure for the RFI.  In essence, (22) and (25) constitute the necessary and sufficient conditions for casting the generalized RFI-Euler theorem (23) in the form specified by (15).  Further, (22), (23), and (25) form the basis for establishing the conjugate relationship between the Lagrange multipliers ($\lambda_i$'s) and the expectation values ($
\left\langle {A_i \left( x \right)} \right\rangle$'s).  \textit{The above results demonstrate the translation of the entire mathematical structure of thermodynamics into the RFI framework}.  Further, the role of the generalized RFI-Euler theorem (23) in deriving the Legendre transform structure is established.

\section{Inference of energy eigenvalues}
The reciprocity relations and Legendre transform structure of the FIM [22] have only employed the data driven information theoretic term $
\left\langle {A_i \left( x \right)} \right\rangle$ as the expectation value. The Schr\"{o}dinger-like link for the RFI significantly modifies this scenario by including the derivatives of the convex potential $V(x)$ into the expectation values, since they form an integral part of the potential of the Schr\"{o}dinger-like link for the RFI.  This is evident in Eq's. (13) and (14). It is important to note that while in the FIM model [22], the data driven terms in the pseudo-potential term of the FIM model constitute an \emph{information-theoretic} potential, the RFI pseudo-potential (14) also has a \emph{physical content} owing to the presence of derivatives of $V(x)$ [5].  The leitmotif of this Section is to seamlessly infer the energy eigenvalues of the Schr\"{o}dinger-like link for the RFI employing only the quantum mechanical virial theorem [19] and the RFI Legendre transform structure (derived in Section 6).  This is accomplished without recourse to numerically evaluating the Schr\"{o}dinger-like link for the RFI.  In this context, the work presented in this Section represents a significant qualitative advancement of the analysis in [25].

Multiplying  (13) by 2 and re-arranging the terms yields
\begin{equation}
 - \frac{{d^2 \psi \left( x \right)}}{{dx^2 }} + \tilde U_{RFI} \left( x \right)\psi \left( x \right) = \frac{{\lambda _0 }}{4}\psi \left( x \right),
\end{equation}
where the RFI pseudo-potential (14) is re-defined as
\begin{equation}
\tilde U_{RFI}\left( x \right) = -\frac{1}{4}\left[ \sum\limits_{i = 1}^M {\lambda _i A_i \left( x \right)}   -  {V^2_x \left( x \right)}   + 2V_{xx} \left( x \right) \right].
\end{equation}
Note that (32) is of the form of the usual time independent Schr\"{o}dinger equation having energy eigenvalue $E$, for $
\frac{{\hbar ^2 }}{{2m}} = 1$ and $
\frac{{\lambda _0 }}{4} = E$.
The quantum mechanical virial theorem for Schr\"{o}dinger models is [17]
\begin{equation}
-\int {\psi \left( x \right)\frac{{d^2 \psi \left( x \right)}}{{dx^2 }}} dx = \left\langle {x\frac{{d \tilde U_{RFI} }}{{dx}}} \right\rangle.
 \end{equation}
The RFI pseudo-potential (33) is expressed in terms of its physical and data driven components as
\begin{equation}
\tilde U_{RFI} \left( x \right) = \tilde U_{RFI}^{Physical} \left( x \right) + \tilde U_{RFI}^{Data} \left( x \right),
\end{equation}
where
\begin{equation}
\begin{array}{l}
 \tilde U_{RFI}^{Physical} \left( x \right) = -\frac{1}{4}\left[ {V_x^2 \left( x \right) - 2V_{xx} \left( x \right)} \right], \\
 and, \\
 \tilde U_{RFI}^{Data} \left( x \right) =  - \frac{1}{4}\sum\limits_{i = 1}^M {\lambda _i A_i \left( x \right)}.  \\
 \end{array}
\end{equation}
Multiplying (34) by 4 and invoking (17) and (35) yields
\begin{equation}
I\left[ \psi  \right] = 4\left\langle {x\frac{{d \tilde U_{RFI} \left( x \right)}}{{dx}}} \right\rangle  = 4\left\langle {x\frac{{d \tilde U_{RFI}^{Physical} \left( x \right)}}{{dx}}} \right\rangle  + 4\left\langle {x\frac{{d \tilde U_{RFI}^{Data} \left( x \right)}}{{dx}}} \right\rangle.
\end{equation}
Eq. (37) yields
\begin{equation}
I\left[ \psi  \right] - 4\left\langle {x\frac{{d \tilde U_{RFI}^{Physical} \left( x \right)}}{{dx}}} \right\rangle  = 4\left\langle {x\frac{{d \tilde U_{RFI}^{Data} \left( x \right)}}{{dx}}} \right\rangle.
\end{equation}
Substituting (36) into (38) and invoking (19) yields
\begin{equation}
\Im \left[ {\psi \left| {e^{ - V\left( x \right)/2} } \right.} \right] =  - \sum\limits_{i = 1}^M {\lambda _i \left\langle {x\frac{{dA_i \left( x \right)}}{{dx}}} \right\rangle }.
\end{equation}
Here, (39) requires that the physical pseudo-potential of the Schr\"{o}dinger-like link for the RFI relates to the convex potential $V(x)$ as
\begin{equation}
4\left\langle {x\frac{{d\tilde U_{RFI}^{Physical} \left( x \right)}}{{dx}}} \right\rangle  = 2\left\langle {V_{xx} \left( x \right)} \right\rangle  - \left\langle {V_x^2 \left( x \right)} \right\rangle.
\end{equation}
Comparison of (20) and (39) results in
\begin{equation}
\lambda _0  + \sum\limits_{i = 1}^M {\lambda _i \left\langle {A_i \left( x \right)} \right\rangle }  =  - \sum\limits_{i = 1}^M {\lambda _i \left\langle {x\frac{{dA_i \left( x \right)}}{{dx}}} \right\rangle }.
\end{equation}

The data driven terms in the RFI pseudo-potential may be expressed
as moments of the independent variable, because the powers $x^k$
constitute a basis in Hilbert space. Thus, without loss of
generality one can write
\begin{equation}
\left\langle {A_i \left( x \right)} \right\rangle  = \left\langle {x^k } \right\rangle.
\end{equation}
Invoking now the Legendre transform derived in (28) yields
\begin{equation}
\frac{{\partial \lambda _0 }}{{\partial \lambda _i }} =  - \left\langle {x^k } \right\rangle.
\end{equation}
Substituting (43) into (41) yields the linear PDE
\begin{equation}
\lambda _0  = \sum\limits_{i = 1}^M {\left( {1 + k} \right)\lambda _i \frac{{\partial \lambda _0 }}{{\partial \lambda _i }}}.
\end{equation}

The procedure for the inference of the energy eigenvalues of the Schr\"{o}dinger-like link for the RFI without recourse to the SWE (13) is tacitly encapsulated in  the linear PDE (44).  Solution of (44) and specifying $
\lambda _0  = 4E$ yields the energy eigenvalues of the the Schr\"{o}dinger-like link for the RFI. \emph{This has only been rendered possible by the invoking of Eq. (19), which is the relationship between the RFI and the FIM.  This explicitly demonstrates the immense utility of (19)}.  \emph{Apart from constituting a significant condition in its own right, (44) lays the basis for establishing a comprehensive quantal connection for the RFI.} Further, (44) requires that the RFI Legendre transform structure be specified in the form derived in Section 6.  By induction, this justifies the physical utility of the generalized RFI-Euler theorem (23).
\section{Summary and discussions}
A critical relation between the RFI and the FIM has been derived (Eq. (11) in Section 3).  It has been shown that the mathematical structures  underlying
thermodynamics can be variationally reproduced by recourse to the
RFI.  This is accomplished in Section 5, in which a principled
generalized RFI-Euler theorem was derived, and in Section 6, in
which thermodynamics' Legendre transform structure were expressed
in terms of the RFI.  The qualitatively distinct nature of these
relations within the context of the RFI framework,
\textit{vis-\'{a}-vis} prior studies utilizing the B-G-S model and
FIM model, have been highlighted. Further, the Schr\"{o}dinger-like link for the RFI which forms the basis for deriving
correspondence relations between the extremal solutions of the RFI
and the FIM and/or the Shannon entropy constrained by Lagrange
multipliers has been established in Section 5.

A thermodynamic basis for employing the
RFI as a measure of uncertainty in areas such as statistical
inference and allied disciplines, parallelling the minimum
relative cross entropy principle [26], has been established.
  Furtherance of this objective comprises the focus of ongoing
work. The efficacy of the relation described by (19), which is the manifestation of (11) in the form of probability amplitudes has been highlighted in Section 7.  Herein, (19) (and thus (11)) facilitates a principled procedure to infer the energy eigenvalues of the Schr\"{o}dinger-like link for the RFI (13), with recourse only to the quantum mechanical virial theorem and the Legendre transform structure of the RFI.

Finally, the results of Sections 5 and 7 give rise to an interesting
conjecture concerning the nature of the wave function $\psi(x)$, for
a physical potential $V(x)$, and, the probability densities
$f(x)=\psi^2(x)$ and $g(x)=\exp{[-V(x)]}$  (note that, in one
dimension, the wave function can always be taken to be real
[20]). Here, $\psi(x)$ is the amplitude that extremizes the
RFI. Thus, \textit{prima facie} the RFI associates the pair
$\{\psi, V(x)\}$ in a definite manner that does not, in principle,
explicitly involve the Schr\"{o}dinger wave equation. Specifically, given $V(x)$ and performing a search
within Hilbert space for the function $\psi(x)$ that extremizes the
RFI, it is mathematically feasible to envisage a method of finding
$\psi(x)$, without recourse to the Schr\"{o}dinger-like link for the RFI.

\section*{Acknowledgements}

RCV gratefully acknowledges communications with G. Blower.  This work was initially conceived under the auspices of \textit{RAND-MSR} contract
\textit{CSM-DI $ \ \& $ S-QIT-101155-03-2009}, and performed under the auspices of \textit{NSFC} contract
\textit{111017-01-2013}.

\end{document}